# Comment on "A self-assembled three-dimensional cloak in the visible" in Scientific Reports 3, 2328

Owen D. Miller[1,*], Wenjun Qiu[2], John D. Joannopoulos[2], Steven G. Johnson[1]
[1]*Dept. of Mathematics, Massachusetts Institute of Technology, Cambridge, MA 02139*
[2]*Dept. of Physics, Massachusetts Institute of Technology, Cambridge, MA 02139*
*odmiller@math.mit.edu

Mühlig et. al. propose and fabricate a "cloak" comprised of nano-particles on the surface of a sub-wavelength silica sphere. However, the coating only reduces the scattered fields. This is achieved by increased absorption, such that total extinction increases at all wavelengths. An object creating a large shadow is generally not considered to be cloaked; functionally, in contrast to the relatively few structures that can reduce total extinction, there are many that can reduce scattering alone.

Cloaking is an exciting, relatively new field generating widespread interest for a variety of applications[1,2]. Mühlig and co-workers[3] theoretically and experimentally demonstrate the ability to reduce scattering cross-sections of sub-wavelength spheres by coating them with metallic nano-particles. The authors provide a thought-provoking demonstration of nanoparticle fabrication and scattering control. However, it is important to note that the reduction in *scattering* comes at the expense of *increased absorption*, such that the total extinction actually *increases* over the entire 250-500nm wavelength range.

Terminology is usually imprecise, and the term "cloaking," while typically referring[1,2,4] to reductions in the total *extinction* (scattering + absorption) cross-section, has also occasionally[5,6] been defined as a reduction in the *scattering* cross-section alone. Defining a "cloak" by its extinction disqualifies objects with large shadows, whereas the scattering definition is a simplification for applications in which "observers" have access only to off-axis, far-field measurements. Regardless of the terminology, it is important to delineate the exact functionality. The relevant applications and technologies for structures that reduce scattering but increase absorption are very different from those that neither scatter nor absorb.

The authors[3] cite prior work[7] showing that in the quasi-static limit (dimensions $\ll \lambda$), there are resonances in the scattered field of coated spheres for particular permittivities. It is well-known[8] that in the quasi-static limit, absorption ($C_{abs} \sim V$) is stronger than scattering ($C_{scat} \sim V^2$), indicating this is likely a pathway to a reduction in scattering cross-section, but not necessarily in total extinction cross-section.

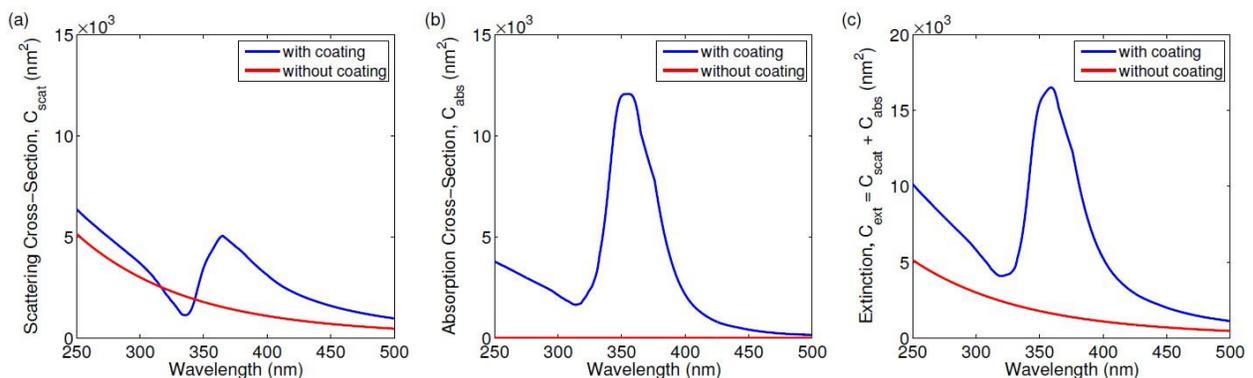

**Figure 1.** (a) Scattering, (b) absorption, and (c) extinction cross-sections for silica shells with and without metal nano-particle coatings. Near $\lambda = 335 nm$, the coating decreases the scattering cross-section. The absorption, however, increases substantially with the coating, and the total extinction – representative of the ``visibility'' of an object – is actually increased by the coating across the entire 250-500nm bandwidth.

We have computed both the scattering and absorption cross-sections for the structures theoretically proposed and fabricated by Mühlig et. al.[3] We used the transfer matrix method[8,9] for the computations, a rigorous full-wave technique for spherically-symmetric structures that is also used by the authors[3] to generate Fig. 5. Fig. 1 depicts the results of the simulations. For the frequencies where the scattering is reducing by the coating, there is also a significant increase in absorption, such that the overall extinction is increased by the coating. At the resonance wavelength of $\lambda \approx 335nm$, the coating increases the apparent size of the object by almost a factor of three. The smallest *enhancement* of the extinction cross-section in the 300-400nm wavelength range is 1.65, such that the coating is almost doubling the apparent size of the object.

There are instances in which one might want to reduce scattering alone, independent of absorption; aircraft obscurance is one example. However, there are a number of well-known structures that can achieve this. As a direct comparison to the work by Mühlig et. al.[3], a 55nm radius silica sphere with a simple 10nm Ag coating can exhibit 3x scattering suppression near 400nm, even though extinction increases by roughly a factor of six. More generally, highly-optimized anti-reflection coatings become the primary competition. Once the requirement to decrease absorption is removed, a variety of well-developed technologies become available and the threshold to make an impact is higher.

**References**

[1] J. B. Pendry, D. Schurig, & D. R. Smith, Controlling electromagnetic fields. *Science* **312**, 1780 (2006)
[2] U. Leonhardt, Optical conformal mapping. *New J. Phys.* **8**, 118 (2006)
[3] S. Mühlig et. al., A self-assembled three-dimensional cloak in the visible. *Scientific Reports* **3**, 2328 (2013)
[4] D. Schurig, J. J. Mock, B. J. Justice, S. A. Cummer, & J. B. Pendry, Metamaterial electromagnetic cloak at microwave frequencies. *Science* **314**, 977 (2006)
[5] A. Alu & N. Engheta, Cloaking a sensor. *Phys. Rev. Lett.* **102**, 233901 (2009)
[6] A. Alu & N. Engheta, Achieving transparency with plasmonic and metamaterial coatings. *Physical Review E* **72**, 016623 (2005)
[7] A. Alu & N. Engheta, Plasmonic and metamaterial cloaking: physical mechanisms and potentials. *J. Opt. A: Pure Appl. Opt.* **10**, 093002 (2008)
[8] C. F. Bohren & D. R. Huffman, *Absorption and Scattering of Light by Small Particles* (John Wiley & Sons, New York, NY, 1983)
[9] W. Qiu et. al., Optimization of broadband optical response of multilayer nanospheres. *Optics Express* **20**, 18494 (2012)